\documentclass[prl,twocolumn,aps,showpacs]{revtex4}
\makeatletter
\def\@pnumwidth{2em}
\makeatother
\usepackage{amssymb,amsmath}
\usepackage[latin2]{inputenc}
\usepackage{graphicx}
\usepackage{psfrag}
\usepackage{dcolumn}
\usepackage{bm}
\def\be{\begin{equation}}
\def\ee{\end{equation}}
\def\ds{\displaystyle}
\def\m2#1#2#3#4{%
\left(
\begin{array}{lr}
\ds{#1}  &  \ds{#2}  \\
\ds{#3}  &  \ds{#4}
\end{array}
\right)
}

\def\dd{D^{(1)}}
\def\du{D^{(2)}}
\def\dt{\delta s}
\def\eps{\varepsilon}
\def\ns{n}

\newcommand{\pd}{\partial}

\begin{document}

\title{Generalized DMPK equation for strongly localized regime - numerical solution}

\author{ J. Brndiar$^1$, R.~Derian$^{1,2}$ and P. Marko\v{s}$^1$}%
\affiliation{$^1$ Institute of Physics, Slovak Academy of
Sciences,  845 11 Bratislava, Slovakia\\
$^2$ Dept. Physics,  Faculty of Electrical Engineering and Information Technology, Ilkovi\v cova 3, 812 19 Bratislava}

\begin{abstract}

Generalized Dorokhov-Mello-Pereyra-Kumar (GDMPK) equation 
[K.~A.~Muttalib and J.~R.~Klauder, Phys. Rev. Lett.  {\bf 82}, 4272 (1999)]
has been proposed for the description of the electron transport in strongly localized 
systems.  We develop an algorithm for the numerical  solution of this equation
and confirm that GDMPK equation correctly describes  the critical and localized regimes.
Contrary to the original DMPK equation, the generalized one contains also an information
about the dimension of the system. In particular, it distinguishes between the two and the
three dimensional models with the same number of transmission channels.
\end{abstract}


\maketitle

Two decades ago, there was the belief that the distribution of the
logarithm of the conductance, $\ln g$,  of the strongly disordered electron system is Gaussian,
independently on the dimension of the system. This paradigm was based on the
two-terminal expression for the conductance,\cite{SE}
\be\label{eco}
g=\ds{\frac{e^2}{h}}~\textrm{Tr}~t^\dag t =\sum_{i=1}^N\ds{\frac{1}{\cosh^2 x_i}}.
\ee
Eq. (\ref{eco}) expresses the conductance in terms of parameters $x_i$ which determine
the eigenvalues of the
matrix $t^\dag t$ ($t$ is the $N\times N$ transmission matrix and 
$N$ is the number of open channels). 
Since all parameters $x_i$ should increase linearly with the system size, $x_i\propto L$,
it is natural to expect that in the limit of large $L$, only the contribution of the
first channel (that one with the smallest parameter $x$) survives and $\ln g \approx -x_1+\ln 4$.
In the analogy with the one-dimensional (1D) case ($N=1$), the distribution $p(\ln g)$ should be Gaussian.
This is  consistent with the solution of  
the  Dorokhov-Mello-Pereyra-Kumar (DMPK) equation, \cite{dmpk} 
for very  long quasi-one dimensional (Q1D) \textsl{weakly disordered} systems.
\cite{Pnato}

However, numerical data \cite{Markos} for \textsl{strongly disordered} 3D samples showed
that $p(\ln g)$ is \textsl{not} Gaussian. Although all $x_i\propto L$, the mean values of the differences
$x_{i+1}-x_i$, $i=1,2,\dots$ are  constant, independent on the size of the system. Therefore,
higher parameters, $x_2$, $x_3$, \dots, affect the form of the distribution $p(\ln g)$ in 3D.
Since the deviation from Gaussian distribution cannot be obtained 
from the  DMPK equation, the generalization of the last has 
been proposed by Muttalib and co-workers.
\cite{mukl,mu-go-02} 

The derivation the DMPK equation\cite{dmpk} is based on 
the parametrization of the transfer matrix ${\bf T}$.
For spin-less electrons and the time reversal symmetry of the system,
\be
{\bf T} = \m2{u}{0}{0}{u^*}\m2{\sqrt{1+\lambda}}{\sqrt{\lambda}}{\sqrt{\lambda}}{\sqrt{1+\lambda}}\m2{v}{0}{0}{v^*},
\ee
where $\lambda$ is a diagonal matrix with diagonal elements $\lambda_i=(1+\cosh 2x_i)/2$,
and $u$, $v$ are unitary matrices.  \cite{pichard} 
For sufficiently long systems, it is assumed that
the elements of matrices $u$, $v$ and $\lambda$ are statistically 
 independent. Contrary to the classical DMPK, the generalized DMPK equation (GDMPK) \cite{mukl} 
 contains additional $N(N-1)$ parameters,
\be
K_{ij}=\left\langle\sum_{a=1}^N |u_{ia}|^2|u_{ja}|^2\right\rangle,
\ee
where $\langle\dots\rangle$ means a mean value  over the realization of the disorder. The GDMPK equation 
reads
\be
\label{gdmpk}
\ell\frac{\pd}{\pd L}P(\{x\})=\ds{\frac{1}{4}} \sum_{i=1}^{N}\frac{\pd}{\pd x_{i}}K_{ii}\left(\frac{\pd P}{\pd x_{i}}+P\frac{\pd}{\pd x_{i}}\Omega(\{x_{n}\})\right).
\ee
Here, $P(\{x\})$ is the common probability distribution of all parameters $x$, $\ell$
is  the mean free path, $\ell$. 
The interaction potential $\Omega(\{x\})$ reads
\be\label{gomega}
\begin{array}{ll}
\Omega(\{x_{n}\})=&-\sum_{i<j}\gamma_{ji}\ln|\sinh^{2}(x_{j})-\sinh^{2}(x_{i})|\\
~~~  &  ~~~\\
~~&-\sum_{i=1}^{N}\ln|\sinh(2x_{i})|,
\end{array}
\ee
and  $\gamma_{ij}=2K_{ij}/K_{ii}$,

Parameters $K_{ij}$ depend on the strength of the
disorder. For weak disorder, 
$K_{ij}=[1+\delta_{ij}]/(N+1)$,  $\gamma_{ij}\equiv 1$ and 
the GDMPK equation reduces to the classical DMPK equation. \cite{dmpk,been} 
For general disorder we expect that $K$ contains an information about the strength of the
disorder,  the topology and the dimensionality of the disordered system \cite{MMW}. For instance,
\cite{mukl}   $K_{11}\sim 1$ and $K_{12}\sim 1/L$
in the localized 3D regime. 
We  expect that  the transition from the
metallic the insulating regime  is associated with (continuous) change of
parameters $K$ \cite{MMWK,MMW}. 
and  the GDMPK equation with correct choice of
parameters $K$  describes the transport in the metallic,  insulating and even in  the critical regime.
Note, the classical DMPK equation
is not  applicable to strongly disordered systems and
contains no information about the dimension.

In Refs. \cite{MMWK,MMW}, a simplified GDMPK equation was  introduced,
with only two parameters: the diagonal elements   $K_{ii}\equiv K_{11}$ for all $i$,
and the off-diagonal elements $K_{ij}\equiv K_{12}$ for all $i\ne j$. \cite{com}
An approximate solutions of such  GDMPK equation, based on the saddle-point method \cite{MW}
was obtained in Refs. \cite{MMWK,MMW}, and the solution was compared with
numerical data obtained by the transfer matrix method (TM). 
Although  quite satisfactory, agreement between approximate solutions and numerical data was obtained,
a quantitative analysis of the  solution of the GDMPK equation is still missing.
Contrary to the DMPK equation, which is exactly solvable \cite{beenakker}, no exact solution
of the GDMPK equation is known. 
It is therefore highly desirable to solve the GDMPK equation numerically.

In this paper, we present  an  algorithm for the numerical solution of   the GDMPK equation.
Our method is based on the mapping of the GDMPK equation onto the Langevin equation,
which describes the diffusion of $N$ particles interacting with potential $\Omega(\{x\})$,
given by Eq. (\ref{gomega}). Simulating such diffusion, we obtain numerical solution of 
classical DMPK equation and the GDMPK equation. 
These solutions are compared 
 with the numerical TM data for the tight-binding Anderson model,
\be\label{ham}
{\cal H} = W\sum_r \eps_r c^\dag_rc_r+t_\parallel \sum_z c^\dag_rc_{r'}+t_\perp\sum c^\dag_rc_{r'}
\ee
which describes the transport of a single electron on the $d$-dimensional lattice. Random energies
$\eps_r$ have  
 zero mean value and variance $1/12$, $W$ measures the strength of the disorder.
Hopping term between two nearest neighboring sites $r$ and $r'$ is $t_\parallel=1$
in the direction of the propagation and $t_\perp=0.4$ in the perpendicular directions.
For this choice of parameters, there are no  evanescent (closed)
channels in 2D and 3D systems. \cite{acta} In 3D,
model (\ref{ham}) exhibits the disorder induced transition from the metal 
to the insulator at the critical point $W_c\approx 9.2$. \cite{MMW,z}

Our numerical method uses the fact that the   
GDMPK equation (\ref{gdmpk}) is a special  Fokker-Planck diffusion equation
\be\label{fpe}
\frac{\pd}{\pd s}P=\left(-\frac{\pd}{\pd x_i}\dd_i(\{x\})+\frac{\pd^2}{\pd x_i^2}\du_i(\{x\})\right)P
\ee
for the one dimensional diffusion of $N$ particles located at $x$ in ``time'' $s=L/\ell$.
The common probability distribution $P(\{x\},s)$  determines the positions of all
particles at time $s$.
In Eq. (\ref{fpe}), $\dd_i$ and $\du_{i}$ are the drift and the diffusion coefficients. 
The Fokker Planck equation (\ref{fpe}) describes the random process
given by the Langevin equation, \cite{risken}
\be
\label{langevin}
\partial x_i/\partial s = h_i(\{x\},s)+g_{i}(\{x\})\Gamma_{i}(s),
\ee
with random white noise force $\Gamma(s)$,
$\langle\Gamma(s)\rangle=0$ and $\langle\Gamma(s)\Gamma(s')\rangle=2\delta(s-s')$,
and  coefficients
\be
\label{hg}
h_{i}=\dd_i-\sqrt{\du_i}\frac{\pd}{\pd x_i}\sqrt{\du_i}
~~~~\textrm{and}~~~~
g_{i}=\sqrt{\du_i}.
\ee
Comparison of Eqs. (\ref{gdmpk}) with Eq. (\ref{fpe}) gives
\be
\label{drift}
\dd_i=-\frac{K_{ii}}{4}\ \frac{\pd\Omega(\{x_n\})}{\pd x_i},
~~~~\textrm{and}~~~~
\du_{i}=\frac{K_{ii}}{4}.
\ee

\begin{figure}[t!]
\centerline{\includegraphics[clip,width=0.28\textheight,angle=-90]{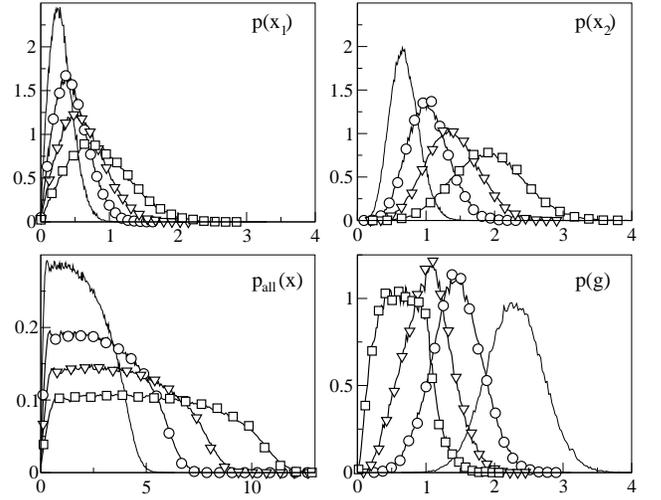}}
\caption{The test of the numerical algorithm. The classical DMPK equation
is solved for $N=9$ channels. Results are compared with the TM data for the Anderson model
on the  lattice $3\times 3\times L$ and disorder $W=2$.  
Solid lines are the  TM data for  $L=30$ (used as initial condition for the DMPK equation), 
and $L=50$,  70,  and 100. 
Symbols show  solutions of the DMPK equation
for  $s=1.8$ (circles),  3.6  (triangles) and 6.35 (squares). For these parameters, both the TM and the DMPK solutions
have the same value of the mean conductance $\langle g\rangle$.
}
\label{fig-test}
\end{figure}

We  simulate the Langevin force $\Gamma$, integrate
the equation of motion (\ref{langevin})  and  take the 
average for a large number of realizations. 
The numerical integration of the equation (\ref{langevin})  gives
\be
\label{nscheme}
x_{i(n+1)}=x_{in}+\dd_i(\{x\})\dt +\sum_{j=1}^{N}\sqrt{\du_i(\{x\})\dt}\times w_{in},
\ee
where $N$ is the number of particles and  $\dt$ is the ``time'' step.
Statistically independent Gaussian variables 
$w_{in}$  have  zero mean and variance 
$
\langle w_{jn}w_{kn'}\rangle=2\delta_{jk}\delta_{nn'}.
$
After $n$ integration steps we obtain the stochastic variables $x_{in}=x_{i}(s)$ at 
``time'' 
$s=\ns\times \dt$.

In simulations, we have to keep the ordering,
 $x_1<x_2<\dots<x_{N}$. Therefore, the position of the $i$th particle is restricted as
 $x_{i-1n}<x_{in}< x_{i+1n-1}$. To keep this constrain, the ``time'' step, $\dt$ must be very small.
For $x_{1n}$ the left boundary is $0$. 
To avoid numerical overflows, we introduce 
and we use the ``cutoff''  $C$,  $x_{Nn}\leqq C,\forall n$. 

\begin{figure}[t!]
\centerline{\includegraphics[clip,width=0.28\textheight,angle=-90]{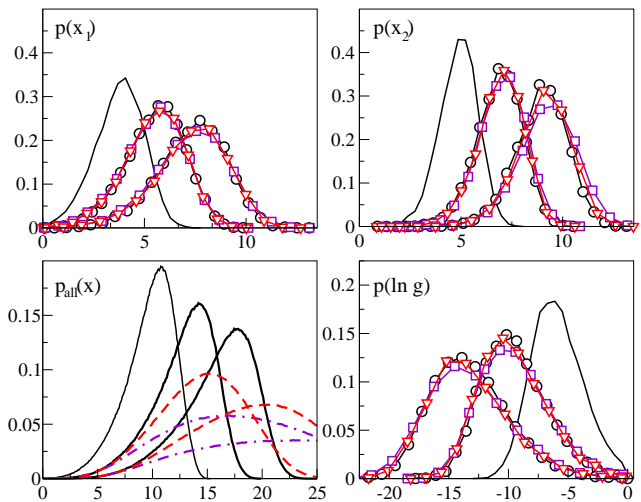}}
\caption{(Color online) The 3D Anderson model $7\times 7\times L$ with disorder $W=29$ (strongly localized regime).
Solid line shows  the TM data for$L=6$, circles are the TM data
for lengths $L=8$ and $L=10$, which are compared with  the GDMPK data  for $s=10.5$ 
and $20 $, respectively. In the  GDMPK equation, we use either entire matrix $K_{ij}$ (triangles)
or the two-parameter model (squares) with $K_{11}=0.5225$ and $K_{12}=0.0244$.
Estimated mean free path $\ell\approx 0.2$ is consistent with \cite{MMW}.
The GDMPK data for $p_{\rm all}(x)$ are shown by
dashed and dot-dashed lines for model with complete matrix  $K$ and the model with 
$K_{11}$ and $K_{12}$, respectively. 
}
\label{49W263D}
\end{figure}

\begin{figure}[t!]
\centerline{\includegraphics[clip,width=0.28\textheight,angle=-90]{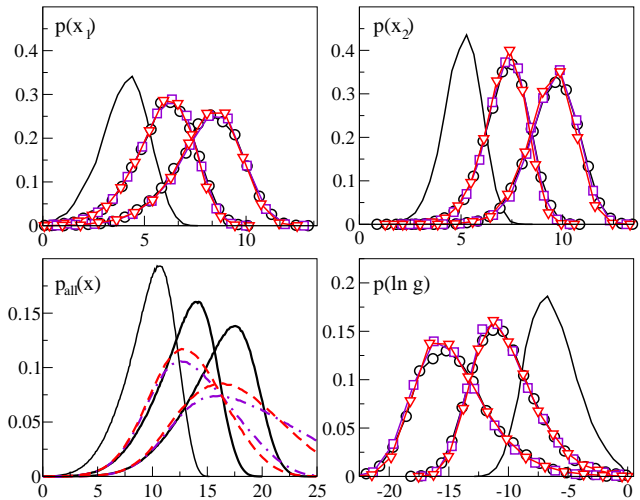}}
\caption{(Color online) The 2D Anderson model $49\times  L$ and $W=26$. Initial condition are given by the 
TM data for $L=6$ (solid line).
The  TM  data for $L=8$ and $L=10$ are  compared with the GDMPK data for  $N=49$ channels and
$s=9.5$ and 19 ($\ell\approx 0.21$).  The  symbols have the same meaning as in Fig. \ref{49W263D}.
We used $K_{11}=0.712$ and $K_{12}=0.0837$ in the two parameter GDMPK equation.
}
\label{49W262D}
\end{figure}

We are not able to start the numerical algorithm 
with the ballistic initial values,
$\lim_{L\rightarrow 0}P=\prod_{i}\delta(\lambda_{i}-0^{+})$,
used in the formulation of the DMPK equation. 
To avoid this difficulty,  we use the TM method and calculate first the
distributions $p(x_i)$ for the Anderson model for a short length $L_0$ of the system.
We use these distribution
as initial values for the GDMPK equation. 
The TM data for $L=L_0$ provides us also  with the values of parameters $K$. 
The solution of the GDMPK equation is compared with the
TM result for length $L>L_0$. In the TM method, the length is defined as a number 
of lattice sites along the propagation direction. For a given $L$, we find  the length $s$  
in the GDMPK equation from the condition that 
 the mean conductance (or the mean of logarithm of the conductance in strongly localized regime) 
found by the two methods  coincides.
We  find in this way also the mean free path (in units of the lattice period  of the Anderson model): 
$\ell=(L-L_0)/s$.

Our results are summarized in Figs. 1-5. 
Compared are four distributions: $p(x_1)$ and $p(x_2)$  for
the two smallest parameters $x$, the distribution  $p_{\rm all}(x)$  of all parameters $x$,
$p_{\rm all}(x)=\langle\sum_i^N \delta(x-x_i)\rangle$, and
the distribution of the conductance $p(g)$  or of the logarithm of the conductance, $p(\ln g)$.

In order   test the numerical algorithm, we solve first the classical DMPK equation.
Fig. \ref{fig-test} shows the data for the weakly disordered Q1D systems with $N=9$.
The length of the system varies between $L=30$ (initial condition for the DMPK) and $L=100$.
The ``time'' step used in the solution of the DMPK equation $\dt=10^{-4}$
and 63.500 iterations were performed. With the use of 
appropriate length $s$, we obtained perfect agreement between the DMPK and TM data, and also
estimate  the mean free path, $\ell\approx 11$, consistent with $\ell\approx 9.2$
estimated for the 3D anisotropic systems \cite{MMW}.

In Figs. \ref{49W263D} and \ref{49W262D}  
we present the solution of  the GDMPK equation for $N=49$ channels in 
the strongly disordered regime (disorder $W=26$).
In Fig. \ref{49W263D} we simulate the transport through the 3D system $7\times 7\times L$.
The TM data for $L=6$ provides us with the initial distributions for the GDMPK as well as with
values of parameters $K$.
Data for $L=8$ and $L=10$ are then compared with two solutions of the GDMPK: in the first simulations,
we use  the entire matrix $K_{ij}$, and  in the second simulation, we substitute all diagonal elements  
by $K_{11}$ and all off-diagonal terms by $K_{12}$. 
Fig. \ref{49W262D} summarizes the TM results for the transport through the 2D strongly disordered 
system $49\times L$ ($W=26$)
compared with the corresponding solution of the GDMPK equation. 

Both Figures show that the GDMPK data for $p(x_1)$, $p(x_2)$ and $p(\ln g)$   agree
with results of the TM simulations. Since the TM data were obtained for  systems of different dimension,
we conclude that the GDMPK equations distinguishes between the 2D and 3D disordered systems.
The information about dimension is  given in parameters $K_{11}$ and $K_{12}$.

\begin{figure}[t!]
\centerline{\includegraphics[clip,width=0.28\textheight,angle=-90]{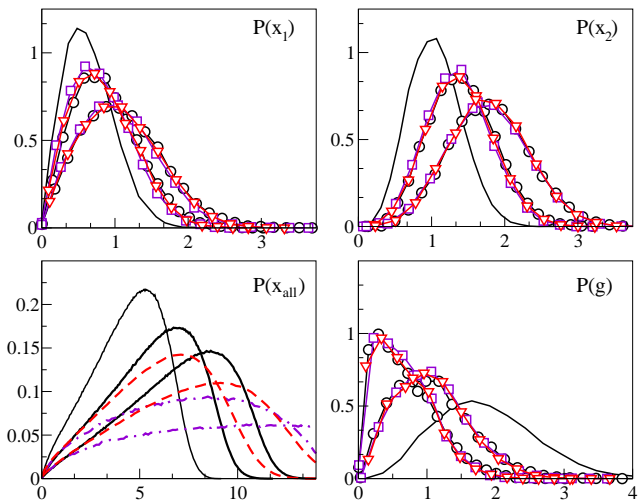}}
\caption{(Color online) The 3D Anderson model $7\times 7\times  L$ with critical disorder $W_c=9.2$ 
Solid lines show the TM data for $L=6$ (initial condition).   The TM data for 
$L=10$ are compared with the  GDMPK data for $N=49$ and  length
$s = 3.25$ and 6.375 ($\ell\approx 0.63$). 
}
\label{49WC}
\end{figure}

\begin{figure}[t!]
\centerline{\includegraphics[clip,width=0.28\textheight,angle=-90]{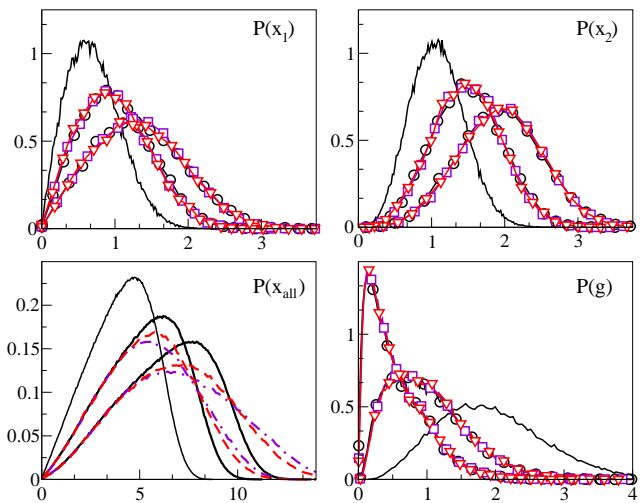}}
\caption{(Color online) The 2D Anderson model $49\times  L$ with  disorder $W=8.6$ 
Solid lines: the TM data for $L=6$ (Initial condition). Circles: the TM data for 
$L=8$ and $L=10$ compared with the GDMPK data for $s=3.375$ and $s=6.5$, respectively
($\ell\approx 0.615$). Symbols have the same meaning as in Fig. 2.
}
\label{492D}
\end{figure}

The difference between the 2D and 3D systems  is  more clearly visible in Figs. \ref{49WC} and \ref{492D}
which compares  3D TM data for the critical disorder $W_c=9.2$
and 2D data for $W=8.6$. Although both system have approximately the same mean free path,
the  conductance  distributions differ  from each other when $L$ increases.

Obtained results confirm that already the simplified two-parameter
GDMPK equation  determines correctly  the statistics of the conductance,
both in the localized and critical regimes.  However,
the total distribution of all parameters $x$, $p_{\rm all}(x)$, is reproduced only
for rather small values of $x$. 
The possible reason for this discrepancy is that 
the TM data for
the entire matrix $K$ might not be sufficiently accurate for high indices $i$ and $j$,
so that we do not know true values of repulsion constants $\gamma_{ij}$ for higher channels.
In  the two parameter model,
$K_{11}$ and $K_{12}$, might overestimate the repulsion of higher channels
(for instance $\gamma_{1i}\sim i^{-1/2}$   \cite{MMW}).
This overestimation  broads the  distribution $p_{\rm all}(x)$.
Interestingly, the 2D data for $p_{\rm all}(x)$ are more  accurate  than the data for 3D.
Nevertheless, as argued previously \cite{jpa}, and confirmed by our data shown in Figs. 2-5,
we do not need to know the distribution of all parameters $x$  for the 
description of the statistics of the conductance.  Only small portion of channels 
is important for the  description of the critical regime in 3D.

In conclusion, we presented numerical solution of the generalized DMPK equation. Our results
confirm  that the generalized DMPK equation describes correctly the electron transport 
in the  localized and the critical regime. The information about the dimension of the system is
given by the additional parameters, $K_{11}$ and $K_{12}$ \cite{com}.
We hope that analytical solution of the generalized DMPK equation could provide us with the analytical description of
the electron transport in the critical and localized regimes in various dimensions.

Acknowledgments. This work was supported by Slovak Grant Agencies APVV, project n. 51-003505, and VEGA, project n. 2/6069/26.


\begin{thebibliography}{99}
\bibitem{com} As was discussed in Ref. \cite{MMW}, the introduction of two additional parameters
$K_{11}$ and $K_{12}$  does not break the validity of the one parameters scaling theory of the transport. 
Indeed, these parameters are the unambiguous functions
of $x_1$ in the limit of $L\to\infty$.
\end{thebibliography}
\end{document}